\documentclass[aps, reprint, superscriptaddress, nofootinbib]{revtex4-1}
\usepackage{amsmath, latexsym, amssymb, hyperref, graphicx, color, slashed}
\definecolor{nicered}{rgb}{0.7,0.1,0.1}
\definecolor{nicegreen}{rgb}{0.1,0.5,0.1}
\hypersetup{colorlinks, citecolor=nicegreen,linkcolor=nicered}

\begin{document}

\newcommand{\Sec}[1]{ \medskip \noindent {\sl \bfseries #1}}
\newcommand{\subsec}[1]{ \medskip \noindent {\sl \bfseries #1}}
\newcommand{\Par}[1]{ \medskip \noindent {\em #1}}

\title{Left-Right Symmetry: from Majorana to Dirac}

\author{Miha Nemev\v{s}ek}
\affiliation{ICTP, Trieste, Italy}
\affiliation{Jo\v{z}ef Stefan Institute, Ljubljana, Slovenia}

\author{Goran Senjanovi\'{c}}
\affiliation{ICTP, Trieste, Italy}

\author{Vladimir Tello}
\affiliation{SISSA, Trieste, Italy}
\affiliation{Arnold Sommerfeld Center, LMU, M\"unchen, Germany}

\date{\today}

\begin{abstract}
Probing the origin of neutrino mass by disentangling the seesaw mechanism is one of the central issues of particle physics.  We address it in the minimal left-right symmetric model and show how the knowledge of light and heavy neutrino masses and mixings suffices to determine their Dirac Yukawa couplings. This in turn allows one to make predictions for a number of high and low energy phenomena, such as decays of heavy neutrinos, neutrinoless double beta decay, electric dipole moments of charged leptons and neutrino transition moments. We also discuss a way of reconstructing the neutrino Dirac Yukawa couplings at colliders such as the LHC.
\end{abstract}

\maketitle

%
%
\Sec{I. Introduction.} 
In the Standard Model (SM) all particles get their masses from the vacuum. This profound mechanism can be verified through the decays of the Higgs-Weinberg boson~\cite{Higgs:1964ia,Weinberg:1967tq}, apparently found by CMS and ATLAS~\cite{CMSATLAS}. In particular, to every charged fermion of mass $m_f$ corresponds a unique (Dirac) Yukawa coupling, which implies the following branching ratio
\begin{equation}\label{hdecay}
  \Gamma \left(h \to f \overline f \right) \propto m_f^2 .
\end{equation}

What about neutrinos? Being neutral, they could be described by real Majorana spinors of mass $m_{\nu}$~\cite{Majorana:1937vz}. This happens naturally in the seesaw mechanism  when one adds heavy right-handed (RH) neutrinos of mass $m_N$ to the SM~\cite{numassreview}. However, even if one were able to measure both light and heavy neutrino masses and light neutrino mixing matrix $V_L$, the Dirac couplings still could not be unambiguously determined~\cite{Casas:2001sr,deGouvea:2010iv}.
This is evident from the expression for the neutrino Dirac yukawa couplings:
\begin{equation} \label{casas}
  M_D =  i \sqrt{m_N} O \sqrt{m_\nu} V_L^{\dagger},
\end{equation}
where $O$ is an arbitrary orthogonal complex matrix. Thus no prediction analogous to~\eqref{hdecay} can be made for neutrinos. The portion of parameter space where the imaginary components of the Euler angles parametrizing $O$ are large, leads to large $\nu-N$ mixing and the origin of neutrino mass is hidden from the processes that could probe it.
 
The question is what happens in a more fundamental theory, such as the left-right (LR) symmetric model, introduced in order to understand the origin of parity violation~\cite{PatiSalam}. Historically, this model led to neutrino masses long before the experiment and also to the seesaw mechanism~\cite{Minkowski, seesaw}.

We show that once the mass matrix of heavy neutrinos is measured, the relation between heavy and light neutrinos can be made definite in the usual manner: one first measures the particle masses and mixing before predicting Yukawa couplings. The~KS~\cite{Keung:1983uu} production process of heavy neutrinos allows one to measure their masses and flavour composition and determine their Majorana nature~\cite{Senjanovic:2010nq}. The theory then predicts the Dirac Yukawa couplings which can in principle be measured at the LHC.  This amounts to probing the origin of neutrino mass, in complete analogy with the Higgs-Weinberg program for the charged fermions and gauge bosons.Moreover, it sheds light on neutrinoless double beta decay and lepton dipole moments. 

%
%
\Sec{II. The Minimal LR Model.}
The minimal left-right symmetric model (LRSM) is based on the gauge group $SU(2)_L \otimes SU(2)_R \otimes U(1)_{B-L}$, augmented by a LR symmetry which implies equality of gauge couplings $g_L = g_R \equiv g$. Fermions come in LR symmetric doublet representations $Q_{L,R} = (u,d)_{L,R}$ and $L_{L,R} = (\nu, \ell)_{L,R}$ and the relevant charged gauge interactions are
\begin{equation} \label{eqWLWR}
  \mathcal L_{gauge} =  \frac{g}{\sqrt 2} \left( 
  \overline \nu_L V_{L}^\dag \slashed{W}_{\!L} e_L + 
  \overline N_R  V_{R}^\dag \slashed{W}_{\!R} e_R\right) + \text{h.c.}.
\end{equation}
The Higgs sector consists~\cite{Minkowski} of a complex bi-doublet $\Phi(2,2,0)$ and two triplets $\Delta_L(3,1,2)$ and $\Delta_R(1,3,2)$ with quantum numbers referring to the LR gauge group. 

In the seesaw picture the Majorana neutrino mass matrix is given by~\cite{MohSenj81}
\begin{equation} \label{eqSeesaw}
  M_\nu = M_L - M_D^T \frac{1}{M_N} M_D,
\end{equation}
where $M_D$ is the neutrino Dirac mass matrix, while $M_L\propto M_{W_L}^2/M_{W_R}$  and $M_N\propto M_{W_R}$ are the symmetric Majorana mass matrices of left- and right-handed neutrinos, respectively. The above formula connects the smallness of neutrino mass to the scale of parity restoration at high energies.

It is crucial that there be new physical phenomena that allow to probe directly\footnote{In case RH neutrinos are too light to be probed at the LHC, one may still determine indirectly their masses and mixings as in the case when the lightest one is the warm dark matter~\cite{Nemevsek:2012cd}.} the Majorana nature of RH neutrinos and determine their masses and mixings from experiment~\cite{Keung:1983uu}, as discussed in the following section.

We opt for charge conjugation $\mathcal C$ as LR symmetry, with the fields transforming as $f_L \leftrightarrow (f_R)^c$, $\Phi \to \Phi^{\text{T}}$ and $\Delta_L \leftrightarrow \Delta_R^*$ (the case of parity will be discussed elsewhere). The mass matrices then satisfy
%
\begin{align} \label{eqDiracSym}
M_L&=\frac{v_L}{v_R} M_N, \\[5pt]
	M_D & = M_D^T,
\end{align}
 where $v_R \equiv \langle \Delta_R^0 \rangle$ sets the large scale (e.g.: $M_{W_R} = g \, v_R$) and $v_L \equiv \langle \Delta_L^0 \rangle$ is naturally suppressed by the large scale and can be shown that $v_L \leq \mathcal{O}(10\text{ GeV})$~\cite{TypeIIRoadmap}. For the complex issues related to determining $v_L$, we refer the reader to~\cite{Perez:2008ha}.

In the case of $\mathcal C$, there is a theoretical lower bound on the LR scale  $M_{W_R} \gtrsim 2.5 \text{ TeV}$~\cite{Maiezza:2010ic, OtherLR}, coming essentially from $K-\overline K$ mixing. It is noteworthy that direct searches for $W_R$ at LHC are now probing this scale~\cite{Nemevsek:2011hz, CMSWR}.

%
%
\Sec{III. From Majorana to Dirac.}
The above seesaw formula seemingly obfuscates the connection between heavy and light neutrinos and common lore was that this connection cannot be unravelled~\cite{Casas:2001sr}. However, since the Dirac mass matrix must be symmetric, it can be obtained directly from~\eqref{eqSeesaw}
\begin{equation} \label{eqMD}
  M_D = M_N \sqrt{\frac{v_L}{v_R} - \frac{1}{M_N} M_\nu},
\end{equation}
and thereby one can determine the mixing between light and heavy neutrinos. The square root of an $n$-dimensional matrix always has $2^n$ discrete solutions which can be found in~\cite{Gantmacher} (ambiguities might arise in singular points of the parameter space). 

The above expression offers a unified picture of the low energy phenomena such as lepton flavour violation, lepton number violation through the neutrinoless double beta decay, electric dipole moments of charged leptons, neutrino transition moments, neutrino oscillations and neutrino cosmology. Some examples are discussed below, while the rest will be dealt with in a forthcoming publication.

It should be mentioned that the determination of the RH neutrino mass matrix as a function of the Dirac Yukawa coupling was studied before in \cite{Falcone:2003af, AkhmedovFrigerio}. This approach requires additional theoretical structure such as quark lepton symmetry and $SO(10)$ unified theories \cite{AkhmedovFrigerio}.   

Here we wish to show,  on the contrary, that without  any new assumption the LRSM is a complete theory of neutrino masses and mixings, in the sense that the measurements of the heavy sector at colliders can determine and inter-connect the low energy phenomena, including those which proceed via Dirac Yukawa couplings. Thus our program is in the same spirit as the SM: to  predict the couplings with the Higgs-Weinberg boson as a function of the basic fermion properties such as masses and gauge mixings. It may take a long time before these Dirac Yukawa couplings are measured; the essential point is the capacity of the theory to relate them to the basic measurable quantities.

\subsec{On the absence of ambiguity of $\boldsymbol{M_D}$.}
As expressed in \eqref{casas}, in the conventional seesaw mechanism $M_D$ is undetermined. On the other hand, in this case (equivalent to setting $v_L=0$ in \eqref{eqMD}), one gets
\begin{equation}\label{typeI}
M_D=i M_N\sqrt{M_N^{-1} M_{\nu}}.
\end{equation}
The crucial point here is that $M_D$ is symmetric and from this requirement the matrix $O$ can be shown to be fixed in terms of  physical parameters $m_{\nu},m_N,V_L$ and $V_R$ (unlike in the case of seesaw in the SM, $V_R$ is a physical parameter as defined in \eqref{eqWLWR})
\begin{equation}\label{fixed-O}
O=\sqrt{m_N}\sqrt{m_N^{-1}V_R^{\dagger}V_L^* m_{\nu} V_L^{\dagger}V_R^{*} }\,V_R^T V_L \sqrt{m_{\nu}^{-1}}.
\end{equation}
As can be seen from above, the elements of $O$ take at most values of order one. Moreover, this parametrisation offers an alternative method of computing $M_D$ which will be discussed elsewhere.
 
The case with nonzero $v_L$ is completely analogous (see \cite{ Akhmedov:2008tb}) and similarly, the matrix $O$ is a function of physical observables only.  

\subsec{$\boldsymbol{M_N}$ from LHC.}
The mass matrix of light neutrinos 
\begin{equation}
  M_\nu = V_L^* m_\nu V_L^\dagger
\end{equation}
is being probed by low energy experiments, while the one of heavy neutrinos\footnote{The mass matrix of charged leptons, being symmetric, can be taken diagonal without loss of generality.}
\begin{equation}
  M_N = V_R m_N V_R^T
\end{equation}
on the other hand, can be determined at high energy colliders through the KS reaction~\cite{Keung:1983uu}. This amounts to producing $W_R$ at the usual Drell-Yan resonance, with a reach of $5.8 \text{ TeV}$ for $W_R$ mass and $3.4 \text{ TeV}$ for the $N$ mass at the LHC~\cite{Ferrari:2000sp, Ginenko:2007}. One can also verify the chirality of the new charged gauge boson~\cite{Ferrari:2000sp, Han:2012}. Unlike in the case of $W_L$, where neutrinos act as missing energy, here the decays of heavy RH neutrinos lead to a lepton number violating final state of two same-sign leptons and two jets. Moreover, one can directly probe the Majorana nature of RH neutrinos through their equal branching ratios into charged leptons and anti-leptons~\cite{Keung:1983uu}. Due to the absence of missing energy in the final state, one can fully reconstruct the heavy neutrino masses $m_N$ from the invariant mass of one of the leptons and two jets in the final state~\cite{Maiezza:2010ic, Nemevsek:2011hz}, together with mixings $V_R$ by tagging the flavour of the final state leptons~\cite{HeavyN}.

While waiting for the LHC to provide this information, the reader may find it useful to have a simple working example
\begin{equation} \label{eqex}
V_R = V_L^*. 
\end{equation}
Although in general~\eqref{eqMD} may require some computational tedium, for this example one gets
 \begin{equation} \label{eqMDs}
  M_D = V_L^* m_N \sqrt{\frac{v_L}{v_R} - \frac{m_\nu}{m_N}} V_L^\dagger.
\end{equation}
It is easy to see from  the generalisation of \eqref{fixed-O} that $O=1$ in this case.

%
%
\Sec{IV. Phenomenological implications.} 
Low scale LRSM contains a host of experimentally accessible phenomena related to lepton number and flavour violation~\cite{Cirigliano:2004tc}, both at high and low energies, which we discuss in this section.

\begin{figure}
	\includegraphics[width=6.6cm]{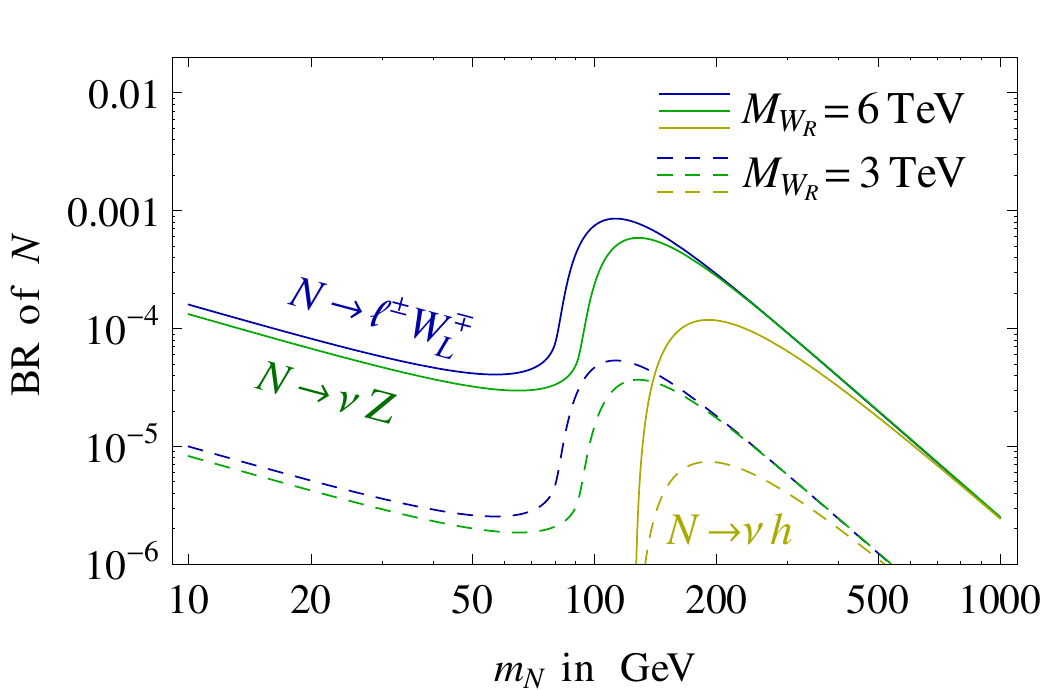}
        \vspace*{-1ex}
	\caption{Branching ratio for the decay of heavy $N$ to the Higgs-Weinberg and SM gauge bosons, proceeding via Dirac couplings, exemplified $v_L = 0$ and $V_R = V_L^*$. The solid (dashed) line corresponds to $M_{W_R} = 6(3) \text{ TeV}$.}
        \vspace{-2ex}
        \label{figNDecay}
\end{figure}

\subsec{$\boldsymbol{N}$ decay at the LHC.}
We start with the high energy probe of $M_D$ at the LHC.
The crucial thing is that $N$, besides decaying through virtual $W_R$ as discussed above, decays also into the left-handed charged lepton through $M_D/M_N$.
In a physically interesting case when $N$ is heavier than $W_L $, which facilitates its search through the KS process, the decay into left-handed leptons proceeds through the on-shell production of $W_L$. For the sake of illustration we choose again the example of \eqref{eqex}, in which case one can estimate the ratio of $N$ decays in the $W_L$ and $W_R$ channels
\begin{equation}
	\frac{\Gamma_{N \to \ell_L j j}}{\Gamma_{N \to \ell_R j j}} \simeq 10^3 \frac{M_{W_R}^4}{M_{W_L}^2 m_N^2} \left| \frac{v_L}{v_R} - \frac{m_\nu}{m_N} \right|,
\end{equation}
which is about a permil for naturally small $v_L$. The branching ratios for the Higgs-Weinberg and SM gauge bosons are shown in Fig.~\ref{figNDecay} (the SM bosons $W, Z, h$ can decay into a lighter $N$, but the small couplings make the corresponding branching ratios too tiny to matter at this point).

The issue here is how to observe these rare channels. Ideally, one should measure the chirality of the outgoing charged lepton~\cite{Ferrari:2000sp, Han:2012} and/or establish the kinematics of the two jets associated with the on-shell production of $W_L$.
This may be a long shot, but could still be feasible for the LHC with a luminosity in the hundreds of $\text{fb}^{-1}$. The bottom line is that this probes in principle all the matrix elements of $M_D$, once the heavy neutrinos are identified through their dominant $W_R$ mediated decays. This offers a clear program of bringing the issue of the origin of neutrino mass to the same level of other fermion masses in the SM.

\begin{figure}
	\includegraphics[width=4.27cm]{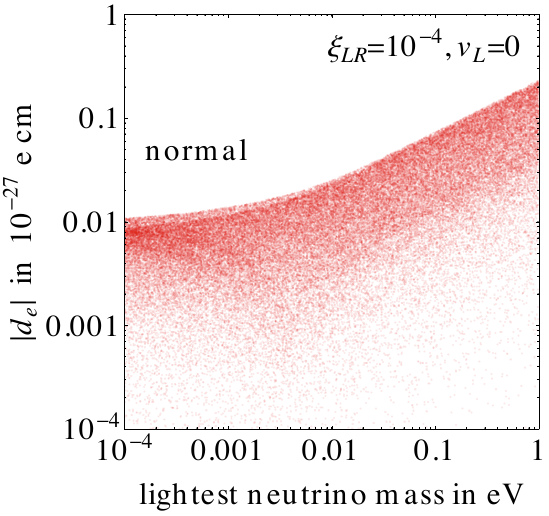}
	\includegraphics[width=4.27cm]{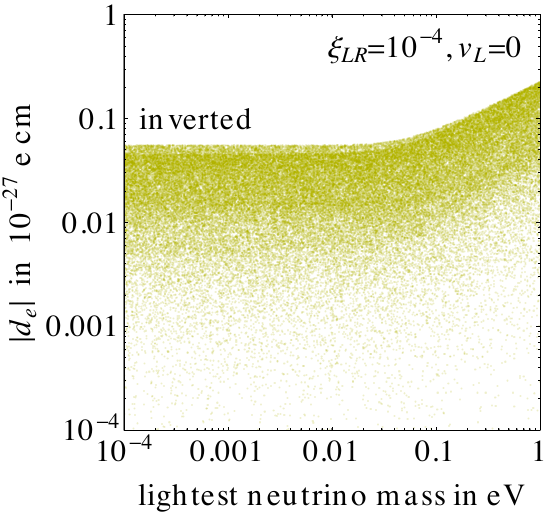}
        \vspace*{-1ex}
	\caption{Electron EDM size in the LRSM with Eqs.~\eqref{eqex}, $v_L=0$ and $m_{N_{1,2,3}} = 0.5, 2, 2.5 \text{ TeV}$. The neutrino mixing angles are fixed at central values provided in~\cite{GonzalezGarcia:2012sz} and the CP phases are scanned over.}
        \vspace{-2ex}
        \label{figEDMPlot}
\end{figure}

\subsec{Electron EDM.}
One of the most sensitive probes of new physics beyond the SM is the T and CP-violating electric dipole moment (EDM) of charged leptons. The SM contribution arises at four loops~\cite{Pospelov:2005pr} and is around eleven orders of magnitude below the current experimental limit $d_e < 10^{-27} \text{ e cm}$~\cite{Hudson:2011zz}. In the LRSM this process is significantly enhanced due to the mixing $\xi_{LR}$ of left and right gauge bosons. The leading amplitude is present at one loop~\cite{Nieves:1986uk, Lavoura:2003xp}  
\begin{equation} \label{eqEDM}
  d_e =  \frac{e G_F }{4\sqrt{2}\pi^2} \text{Im}\left[\xi_{LR} V_R F(t) V_R^{\dagger} M_D \right]_{ee},
\end{equation}
where
\begin{equation}
  F(t) = \frac{t^2-11t+4}{2(t-1)^2}+\frac{3t^2\log{t}}{(t-1)^3},\,  ~ ~t=\frac{m_N^2}{M_{W_L}^2}.
\end{equation}
There are strong limits on the $\xi_{LR}$ but, in any case, it is automatically small due to the suppression of the heavy gauge boson mass. It is bounded by
\begin{equation}
 Ê\frac{\alpha}{4 \pi} \frac{m_t m_b}{M_{W_R}^2} \lesssim \left| \xi_{LR} \right| \lesssim \frac{M_{W_L}^2}{M_{W_R}^2},
\end{equation}
with a lower bound resulting from radiative electroweak corrections~\cite{Branco:1978bz}.

Taking the example in~\eqref{eqex}, the size of the EDM is shown in Fig.~\ref{figEDMPlot} as a function of the lightest neutrino mass for two different neutrino hierarchies. These values can be probed by future experiments~\cite{futureEDM}. In the case when the LR mixing is close to its lower bound, one has to go beyond the one loop approximation~\cite{Atwood:1990tu}, but in that case the experimental outlook seems bleak and we do not pursue it here.

In the context of LRSM, EDM is a manifestly CP-odd process sensitive to Majorana and Dirac phases, complementary to~\cite{deGouvea:2002gf}. This can easily be checked using the example of Eq.~\eqref{eqMDs} in the EDM expression in Eq.~\eqref{eqEDM} where the CP phases do not cancel out.

\subsec{Neutrinoless double beta decay.}
The importance of this textbook example of lepton number violation was recognized in~\cite{Racah:1937qq} soon after the seminal work of Majorana~\cite{Majorana:1937vz}. The LRSM offers new sources  for this process~\cite{MohSenj81} that has been studied extensively over the years~\cite{Rodejohann:2011mu}. In particular, an in-depth analysis~\cite{Tello:2010am} (see also~\cite{Chakrabortty:2012mh}) was recently performed on the connection between neutrinoless double beta decay  (and lepton flavour violation) at low energies  and the KS process~\cite{Keung:1983uu} at colliders.

Although the main source of this process in LRSM is due to the exchange of the heavy neutrinos, there is an additional contribution proportional to the Dirac mass matrix. We express it in the usual form of  an effective mass term
\begin{equation} \label{eqmnuNee}
  m^{ee}_{\nu N} = \left(\xi_{LR} + \eta \frac{M_{W_L}^2}{M_{W_R}^2} \right) p \left(M_N^{-1} M_D \right)_{ee},
\end{equation}
where $p\simeq 100$ MeV~\cite{Tello:2010am} and $\eta\simeq 10^{-2}$ \cite{Doi:1985dx} are determined by nuclear physics considerations.

\begin{figure}
	\includegraphics[width=4.27cm]{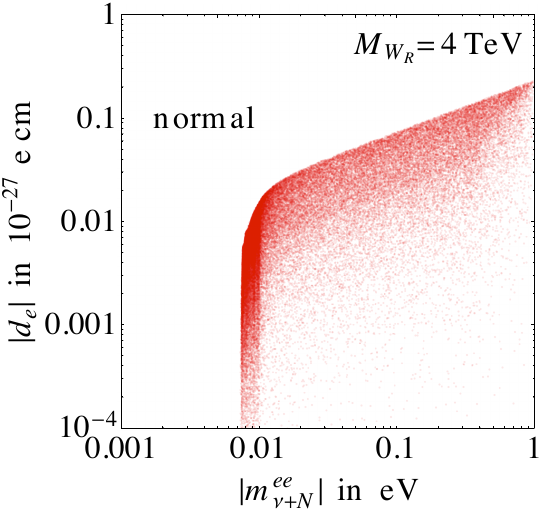}
	\includegraphics[width=4.27cm]{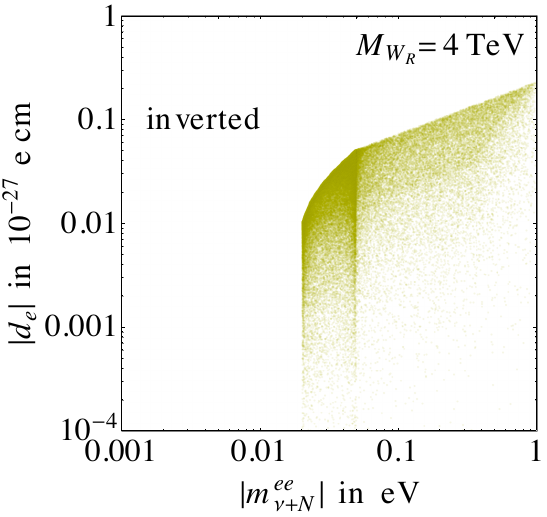}
        \vspace*{-1ex}
	\caption{Electron EDM size and the effective mass for neutrinoless double beta computed with the same choice of parameters as in Fig.~\ref{figEDMPlot}. The mass of $W_R$ is fixed at 4 TeV and $p = 193 \text{ MeV}$~\cite{Tello:2010am}.}
        \vspace{-2ex}
        \label{figBetaEdmPlot}
\end{figure}

As a consequence of~\eqref{eqMD}, the contribution in~\eqref{eqmnuNee} is subleading for naturally small values of $v_L$ (a possible contribution of heavy neutrinos without $W_R$~\cite{Atre:2009rg} is also suppressed) and the total decay rate is governed by the effective mass parameter
\begin{equation}
	|m_{\nu+N}^{ee}|^2 =  \left|ÊV_{Lej}^2 m_{\nu_j} \right|^2 + \left| p^2 \frac{M_{W_L}^4}{M_{W_R}^4} Ê\frac{V_{Rej}^2}{m_{N_j}} \right|^2 .
\end{equation}
Since $|m_{\nu+N}^{ee}|$ and the size of the electron EDM both depend on the heavy neutrino mass, there is a correlation between the two processes, which is shown in Fig.~\ref{figBetaEdmPlot}. The values of $m_N$ are chosen for illustration purpose only.

\subsec{Neutrino transition moments.}
By defining
\begin{equation}
  M = \xi_{LR}^* V_L^T m_\ell M_N^{-1} M_D V_L,
\end{equation}
we get the following matrix of  neutrino magnetic transition moments
\begin{equation} \label{eqMuLR}
	\mu = \frac{i e G_F}{\sqrt 2 \pi^2} \text{Im}\left[ M + M^\dagger\right].
\end{equation}
This result was already derived in~\cite{Czakon:1998rf} and  can be reproduced using~\cite{Lavoura:2003xp}. Here, we neglect the contribution from light neutrino masses, which is roughly nine orders of magnitude smaller than the current experimental limit $\mu < 2 \times 10^{-10} \mu_B$~\cite{Broggini:2012df}. One should keep in mind that the Majorana transition moments in the SM (with non-zero neutrino mass) are negligibly small $\mu_{SM}\simeq 10^{-23}\mu_{B}$~\cite{Broggini:2012df}. 

It is easy to see that~\eqref{eqMuLR} gives roughly $\mu \simeq 10^{-19} \mu_B$ for generic values of $M_D$ in \eqref{eqMD}, still a hopelessly small value. Therefore an observation of neutrino transition moments  would deal a serious blow to the LRSM.

\Sec{V. Conclusions and Outlook.}
In the SM the knowledge of charged fermion masses uniquely predicts Higgs decay branching ratios. As shown here, exactly the same happens in the LRSM for the masses of light and heavy neutrinos. The reason behind this is the LR symmetry itself
which allows one to compute the Dirac Yukawa couplings in the context of the  seesaw mechanism.

The main result of our paper is summarised in Eq.~\eqref{eqMD}. Its phenomenological impact is exemplified both on high energy frontier at the LHC and on the phenomena of neutrinoless double beta decay, dipole moments of charged leptons and neutrino transition moments. This result was achieved at no expense of imposing additional ad-hoc symmetries but by the structure of the theory itself. The bottom line is that one can predict and measure the Dirac neutrino Yukawa couplings in complete analogy with SM situation for charged fermions.

It is interesting to compare our program to the one followed over the years in the quark sector of the LRSM. Here, we took the conventional path of predicting the Yukawa couplings from the knowledge of particle masses and mixings. In the quark sector, on the contrary, the symmetry of quark mass matrices was historically used to fix the flavour structure of the right-handed gauge interaction, which led to the strict bound on the LR scale. Now, with the advent of LHC, the conventional route can be taken up again.

\Sec{Acknowledgments.} We thank A. Melfo for the inspirational version of the first abstract and F. Nesti for the early stages of collaboration and an important suggestion. We thank B. Bajc, A. Melfo and F. Nesti for a careful reading of the manuscript. Also, we would like to thank K. Babu and Y. Zhang for useful discussions and comments. We acknowledge the continuing warm hospitality of BIAS. Special thanks are due to Ivo and Sonja Brzovi\'c for their kind hospitality, care and encouragement. V.T. is deeply grateful to Deniz for support. Lastly, we very much appreciate the insightful and careful comments of the Referees, which helped improve both the physics and presentation of this work.

\end{document}